\documentclass[a4paper,12pt,dvips,oneside]{article}
\usepackage{booktabs,graphicx,overcite,xspace,amsmath,amssymb,longtable,multirow}
\usepackage[scanall]{psfrag}
\usepackage[small,sc]{caption}
\usepackage[dvips]{color}

\begin{document}
\def\aciee{Angew. Chem. Int. Ed. Engl. }
\def\ac{Acta. Crystallogr. }
\def\acp{Adv. Chem. Phys. }
\def\acr{Acc. Chem. Res. }
\def\ajp{Am. J. Phys. }
\def\ap{Ann. Physik }
\def\apc{Adv. Prot. Chem. }
\def\arpc{Ann. Rev. Phys. Chem. }
\def\cccc{Coll. Czech. Chem. Comm. }
\def\cpc{Comp. Phys. Comm. }
\def\cpl{Chem. Phys. Lett. }
\def\crev{Chem. Rev. }
\def\el{Europhys. Lett. }
\def\ic{Inorg. Chem. }
\def\ijmpc{Int. J. Mod. Phys. C }
\def\ijqc{Int. J. Quant. Chem. }
\def\jcis{J. Colloid Interface Sci. }
\def\jcsft{J. Chem. Soc., Faraday Trans. }
\def\jacs{J. Am. Chem. Soc. }
\def\jas{J. Atmos. Sci. }
\def\jbc{J. Biol. Chem. }
\def\jcc{J. Comp. Chem. }
\def\jcp{J. Chem. Phys. }
\def\jce{J. Chem. Ed. }
\def\jcscc{J. Chem. Soc., Chem. Commun. }
\def\jetp{J. Exp. Theor. Phys. (Russia) }
\def\jmb{J. Mol. Biol. }
\def\jmsp{J. Mol. Spec. }
\def\jmst{J. Mol. Struct. }
\def\jncs{J. Non-Cryst. Solids }
\def\jpa{J. Phys. A }
\def\jpc{J. Phys. Chem. }
\def\jpca{J. Phys. Chem. A }
\def\jpcb{J. Phys. Chem. B }
\def\jpcm{J. Phys. Condensed Matter. }
\def\jpcs{J. Phys. Chem. Solids. }
\def\jpsj{J. Phys. Soc. Jpn. }
\def\jrnist{J. Res. Natl. Inst. Stand. Technol. }
\def\mg{Math. Gazette }
\def\mp{Mol. Phys. }
\def\nat{Nature }
\def\nsb{Nat. Struct. Biol.}
\def\Pa{Physica A }
\def\pac{Pure. Appl. Chem. }
\def\pccp{Phys. Chem. Chem. Phys. }
\def\phys{Physics }
\def\pmb{Philos. Mag. B }
\def\ptrsb{Philos. T. Roy. Soc. B }
\def\pnasu{Proc. Natl. Acad. Sci. USA }
\def\pr{Phys. Rev. }
\def\prep{Phys. Reports }
\def\pra{Phys. Rev. A }
\def\prb{Phys. Rev. B }
\def\prbcm{Phys. Rev. B }
\def\prc{Phys. Rev. C }
\def\prd{Phys. Rev. D }
\def\pre{Phys. Rev. E }
\def\prl{Phys. Rev. Lett. }
\def\prsa{Proc. R. Soc. A }
\def\psfg{Proteins: Struct., Func. and Gen. }
\def\sci{Science }
\def\spj{Sov. Phys. JETP }
\def\ss{Surf. Sci. }
\def\tca{Theor. Chim. Acta }
\def\zpb{Z. Phys. B. }
\def\zpc{Z. Phys. Chem. }
\def\zpd{Z. Phys. D }
\def\aciee{Angew. Chem. Int. Ed. Engl. }
\def\ac{Acta. Crystallogr. }
\def\acp{Adv. Chem. Phys. }
\def\acr{Acc. Chem. Res. }
\def\ajp{Am. J. Phys. }
\def\ap{Adv. Phys. }
\def\arpc{Ann. Rev. Phys. Chem. }
\def\cccc{Coll. Czech. Chem. Comm. }
\def\cpl{Chem. Phys. Lett. }
\def\crev{Chem. Rev. }
\def\dalton{J. Chem. Soc., Dalton Trans. }
\def\el{Europhys. Lett. }
\def\faraday{J. Chem. Soc., Faraday Trans. }
\def\fartrans{J. Chem. Soc., Faraday Trans. }
\def\fdisc{J. Chem. Soc., Faraday Discuss. }
\def\ic{Inorg. Chem. }
\def\ijqc{Int. J. Quant. Chem. }
\def\jcis{J. Colloid Interface Sci. }
\def\jcsft{J. Chem. Soc., Faraday Trans. }
\def\jacs{J. Am. Chem. Soc. }
\def\jas{J. Atmos. Sci. }
\def\jcc{J. Comp. Chem. }
\def\jcp{J. Chem. Phys. }
\def\jce{J. Chem. Ed. }
\def\jcscc{J. Chem. Soc., Chem. Commun. }
\def\jetp{J. Exp. Theor. Phys. (Russia) }
\def\jmc{J. Math. Chem. }
\def\jmsp{J. Mol. Spec. }
\def\jmst{J. Mol. Structure }
\def\jncs{J. Non-Cryst. Solids }
\def\jpc{J. Phys. Chem. }
\def\jpcm{J. Phys. Condensed Matter. }
\def\jpsj{J. Phys. Soc. Jpn. }
\def\jsp{J. Stat. Phys. }
\def\mg{Math. Gazette }
\def\mp{Mol. Phys. }
\def\molphys{Mol. Phys. }
\def\nat{Nature }
\def\pac{Pure. Appl. Chem. }
\def\phys{Physics }
\def\pla{Phys. Lett. A }
\def\plb{Phys. Lett. B }
\def\phm{Philos. Mag. }
\def\pmb{Philos. Mag. B }
\def\pnas{Proc.\ Natl.\ Acad.\ Sci.\  USA }
\def\pr{Phys. Rev. }
\def\pra{Phys. Rev. A }
\def\prb{Phys. Rev. B }
\def\prc{Phys. Rev. C }
\def\prd{Phys. Rev. D }
\def\pre{Phys. Rev. E }
\def\prl{Phys. Rev. Lett. }
\def\prsa{Proc. R. Soc. A }
\def\ss{Surf. Sci. }
\def\sci{Science }
\def\tca{Theor. Chim. Acta }
\def\zpc{Z. Phys. Chem. }
\def\zpd{Z. Phys. D }
\def\zfpd{Z. Phys. D }
\def\zpdamc{Z. Phys. D }
\def\aciee{Angew. Chem. Int. Ed. Engl. }
\def\ac{Acta. Crystallogr. }
\def\acp{Adv. Chem. Phys. }
\def\acr{Acc. Chem. Res. }
\def\ajp{Am. J. Phys. }
\def\am{Adv. Mater. }
\def\apl{Appl. Phys. Lett. }
\def\arpc{Ann. Rev. Phys. Chem. }
\def\mrsb{Mater. Res. Soc. Bull. }
\def\cccc{Coll. Czech. Chem. Comm. }
\def\cj{Comput. J. }
\def\cp{Chem. Phys. }
\def\cpc{Comp. Phys. Comm. }
\def\cpl{Chem. Phys. Lett. }
\def\crev{Chem. Rev. }
\def\el{Europhys. Lett. }
\def\fd{Faraday Disc. }
\def\ic{Inorg. Chem. }
\def\ijmpc{Int. J. Mod. Phys. C }
\def\ijqc{Int. J. Quant. Chem. }
\def\jcis{J. Colloid Interface Sci. }
\def\jcsft{J. Chem. Soc., Faraday Trans. }
\def\jacs{J. Am. Chem. Soc. }
\def\jap{J. Appl. Phys. }
\def\jas{J. Atmos. Sci. }
\def\jcc{J. Comp. Chem. }
\def\jcp{J. Chem. Phys. }
\def\jce{J. Chem. Ed. }
\def\jcscc{J. Chem. Soc., Chem. Commun. }
\def\jetp{J. Exp. Theor. Phys. (Russia) }
\def\jmsp{J. Mol. Spec. }
\def\jmst{J. Mol. Structure }
\def\jncs{J. Non-Cryst. Solids }
\def\jpa{J. Phys. A }
\def\jpc{J. Phys. Chem. }
\def\jpcssp{J. Phys. C: Solid State Phys. }
\def\jpca{J. Phys. Chem. A. }
\def\jpcb{J. Phys. Chem. B. }
\def\jpcm{J. Phys. Condensed Matter. }
\def\jpcs{J. Phys. Chem. Solids. }
\def\jpsj{J. Phys. Soc. Jpn. }
\def\jpfmp{J. Phys. F, Metal Phys. }
\def\mg{Math. Gazette }
\def\mp{Mol. Phys. }
\def\msr{Mater. Sci. Rep. }
\def\nat{Nature }
\def\njc{New J. Chem. }
\def\pac{Pure. Appl. Chem. }
\def\phys{Physics }
\def\pma{Philos. Mag. A }
\def\pmb{Philos. Mag. B }
\def\pml{Philos. Mag. Lett. }
\def\pnasu{Proc. Natl. Acad. Sci. USA }
\def\pr{Phys. Rev. }
\def\prep{Phys. Reports }
\def\pra{Phys. Rev. A }
\def\prb{Phys. Rev. B }
\def\prc{Phys. Rev. C }
\def\prd{Phys. Rev. D }
\def\pre{Phys. Rev. E }
\def\prl{Phys. Rev. Lett. }
\def\prsa{Proc. R. Soc. A }
\def\pss{Phys. State Solidi }
\def\pssb{Phys. State Solidi B }
\def\rmp{Rev. Mod. Phys. }
\def\rpp{Rep. Prog. Phys. }
\def\sci{Science }
\def\ss{Surf. Sci. }
\def\tca{Theor. Chim. Acta }
\def\tetra{Tetrahedron }
\def\zpb{Z. Phys. B. }
\def\zpc{Z. Phys. Chem. }
\def\zpd{Z. Phys. D }

\title{Analysis of cooperativity and localization for atomic rearrangements}
\author{Semen A. Trygubenko\footnote{email: sat39@cam.ac.uk}~~and David J. Wales\footnote{email: dw34@cam.ac.uk} \\
{\it University Chemical Laboratories, Lensfield Road,} \\
{\it Cambridge CB2 1EW, UK} }

\maketitle

\begin{abstract}
We propose new measures of localization and cooperativity for the analysis of atomic rearrangements.
We show that for both clusters and bulk material cooperative rearrangements usually have significantly lower barriers
than uncooperative ones, irrespective of the degree of localization.
We also find that previous methods used to sample stationary points
are biased towards rearrangements of particular types.
Linear interpolation between local minima in double-ended transition state searches
tends to produce cooperative rearrangements, while
random perturbations of all the coordinates, as sometimes used in single-ended searches, has the opposite effect.
\end{abstract}

\section{Introduction}
The potential energy surface (PES) governs the observed structure, dynamics and thermodynamics
of any molecular system. It is often possible to gain new insight into these properties by
expressing them in terms of stationary points of the PES,
i.e.~points where the gradient of the potential vanishes.\cite{Wales03}
The most important stationary points are minima and the transition states that
connect them. Here we define a minimum as a stationary point where the Hessian, the second derivative matrix,
has no negative eigenvalues, while a transition state is
a stationary point with precisely one such eigenvalue.\cite{murrelll68}

The number of stationary points on the PES generally scales exponentially
with system size~\cite{walesd03,doyew02,stillinger99,stillingerw82,stillingerw84},
which necessitates an appropriate sampling strategy of some sort for larger systems.
In particular, to analyse dynamical properties a database of local minima and the transition states
that connect them is usually constructed, which generally involves extensive use of
single-ended and double-ended transition state searching techniques
(see Refs.~\citen{Wales03,trygubenkow04,henkelmanjj00} and references therein).
Single-ended transition state searches only
require an initial starting geometry. However, double-ended searches require two endpoint geometries,
a mechanism to generate a set of configurations between them, and a suitable functional (or gradient)
to be evaluated and minimised. The most successful single- and double-ended methods
currently appear to be based upon hybrid eigenvector-following~\cite{crippens71,pancir74,hilderbrandt77,cerjanm81,
simonsjto83,banerjeeass85,baker86,wales94a,walesu94,munrow97,munrow99,kumedamw01,walesw96} and the
nudged elastic band approach~\cite{henkelmanjj00,jonssonmj98,henkelmanuj00,trygubenkow04,xielw04,petershbc04,crehuetf03},
respectively. The two search types are often used together, since double-ended transition state searches do not
produce a tightly converged transition state and further refinement may be needed.\cite{Wales03,trygubenkow04}

Any path connecting two minima on PES can be broken down into elementary rearrangements, each of which
involves a single transition state. The corresponding mechanism can be analyzed in detail by calculating the two unique
steepest-descent paths that lead downhill from the transition state.

The number of elementary rearrangements, as defined above, increases exponentially with system size
as for the number of transition states. For instance,
there are approximately 30,000 such pathways on the PES of the 13-atom cluster bound by the Lennard-Jones potential.
When permutation-inversion isomers are included, this number increases by a factor of order $2 \times N!$.\cite{Wales03}
Two activation barriers can be defined for each pathway in terms of the energy difference between the transition state
and each of the minima. For non-degenerate rearrangements~\cite{Wales03,leones70} the two sides of the path
are termed uphill and downhill, where the uphill barrier is the larger one, which leads to the higher minimum.
The barriers and the normal modes of the minima and transition states can be used to calculate
rate constants using harmonic transition state theory.\cite{Eyring35,EvansP35,Laidler87}

For each local minimum a catchment basin can be defined in terms of all the configurations from which steepest-descent paths
lead to that minimum.\cite{mezey81} Some of these paths originate from transition states on the boundary of the catchment basin, 
which connect a given minimum to adjacent minima. The integrated path length for such rearrangements provides
a measure of the separation between local minima, and may be related to the density of stationary points in configuration
space. The integrated path length is usually approximated as the sum of Euclidean distances between configurations sampled
along appropriate steepest-descent paths.\cite{Wales03}
It provides a convenient coordinate for monitoring the progress of the reaction.

Calculated pathways can always be further classified mechanistically. For example, some rearrangements
preserve the nearest-neighbour coordination shell for all the atoms. In previous studies of bulk models these
cage-preserving pathways were generally found to outnumber the more localized cage-breaking processes, which are
necessary for atomic transport.\cite{middletonw01}
It was found that the barriers for cage-breaking and cage-preserving processes were similar for bulk LJ systems, while
the cage-breaking mechanisms have significantly higher barriers for bulk silicon modelled by the Stillinger-Weber
potential.\cite{middletonw01}

For minima separated by increasing distances in configuration space, the pathways that connect them are likely
to involve more and more elementary steps, and are not unique. Finding such  paths in high-dimensional systems can become
a challenging task.\cite{Wales02,trygubenkow04} Some difficulties have been 
attributed to instabilities and inefficiencies in transition state searching algorithms,\cite{trygubenkow04,maragakisabrk02}
as well as the existence of very different barrier and path length scales.\cite{trygubenkow04}
A new algorithm for locating multi-step pathways in such cases has recently been proposed.\cite{Wales02,trygubenkow04} 

In the present work we have used a doubly-nudged elastic band (DNEB) method~\cite{trygubenkow04}
in conjunction with eigenvector-following (EF)
algorithms~\cite{crippens71,pancir74,hilderbrandt77,cerjanm81,simonsjto83,banerjeeass85,baker86,wales94a,
walesu94,munrow97,munrow99,kumedamw01,walesw96}
to locate rearrangement pathways in various systems. The Lennard-Jones (LJ) potential, defined as
$V(r) = 4 \epsilon \left[ \left(\sigma/r\right)^{12} - \left(\sigma/r\right)^{6} \right]$, where $r$ is the distance
between two atoms, $\epsilon$ is the depth of the potential energy well, and $2^{1/6}\sigma$ is the pair equilibrium separation,
was used to describe the $13$- and $75$-atom Lennard-Jones clusters, LJ$_{13}$ and LJ$_{75}$.
We have also considered a binary LJ (BLJ) system with parameters $\sigma_{AA}=1.0$, $\sigma_{BB}=0.88$, $\sigma_{AB}=0.8$,
$\epsilon_{AA} = 1.0$, $\epsilon_{BB} = 0.5$, $\epsilon_{AB} = 1.5$, where $A$ and $B$ are atom types.
The mixture with $A:B$ ratio $80:20$ provides a popular model bulk glass-former.\cite{middletonw01,sastryds98}
We employed a periodically repeated cubic cell containing $205$ $A$ atoms and $51$ $B$ atoms.
The density was fixed at $1.2\sigma_{AA}^{-3}$ and the Stoddard-Ford scheme was used to prevent discontinuities.\cite{stoddardf73}

The motivation for this paper was our observation that
construction of some multi-step pathways using the connection algorithm described in Ref.~\citen{trygubenkow04}
is particularly difficult.
We were unable to relate these difficulties to simple properties of the pathways such as the
integrated path length, the uphill and downhill
barriers, or the barrier and path length asymmetries.
Instead, more precise measures of localization and cooperativity are required, as shown in the following sections.
It also seems likely that such tools may prove useful in 
analysing the dynamics of supercooled liquids, where
processes such as intrabasin oscillations and interbasin hopping
have been associated with different rearrangement mechanisms.\cite{vogeldhg04}
In particular, cooperativity is believed to play an important role at low
temperatures in glass-forming systems~\cite{arndtsghk97},
and dynamic heterogeneity may result in decoupling between structural relaxation and
transport properties for supercooled liquids.\cite{berthier04}

\section{Methods}
\label{sec:methods}
\subsection{Localization}
The outcome of a pathway calculation for an atomic system will generally be
a set of intermediate geometries, and the corresponding energies, for points along
the two unique steepest-descent paths that link a transition state to two local
minima.
This discrete representation is a convenient starting point for our analysis of localization and
cooperativity. 
We number the structures along the path $j=1,\ 2,\ldots,\ N_f$ starting from one of the two minima
and reversing the other steepest-descent path, so that structure $N_f$ corresponds to the
other minimum. The transition state then lies somewhere between frames $1$ and $N_f$.
We define the three-dimensional vector ${\bf r}_i(j)$ to contain the 
Cartesian coordinates of atom $i$ for structure $j$, i.e.~${\bf r}= \Bigl(X_i(j),Y_i(j),Z_i(j)\Bigr)$,
where $X_i(j)$ is the $X$ coordinate of atom $i$ in structure $j$, etc.

For each atom $i$ we also define the displacement between structures $j-1$ and $j$ as
\begin{equation}
	d_i(j) = \Bigl| {\bf r}_i(j) - {\bf r}_i(j-1) \Bigr|.
\end{equation}
Hence the sum of displacements
\begin{equation}
	d_i = \sum_{j=2}^{N_f} d_i(j),
	\label{eq:di}
\end{equation}
is an approximation to the integrated path length for atom $i$, which becomes increasingly accurate
for smaller step sizes.
The total integrated path length, $s$, is approximated as
\begin{equation}
     s = \sum_{j=2}^{N_f} \sqrt{\sum_{i=1}^N d_i\left(j\right)^2},
\end{equation}
where $N$ is the total number of atoms. $s$ is a characteristic
property of the complete path, and is expected to correlate with
parameters such as the curvature and barrier height for short paths.\cite{Wales01,Wales03,bogdanw04}

The set $\{d_1,d_2,\ldots,d_N\}$ containing all $N$ values of $d_i$ will be denoted $\{d\}$,
and analogous notation will be used for other sets below.
We will also refer to the frequency distribution
function, which can provide an alternative representation of such data.\cite{bulmer79}
For example, the frequency distribution function $F$
for a given continuous variable, $x$, tells us that
$x$ occurs in a certain interval $F(x)$ times. 

Our objective in the present analysis is to provide a more detailed description of
the degree of `localization' and `cooperativity' corresponding to a given pathway.
The first index we consider is $N_p$, which is designed to provide
an estimate of how many atoms {\em participate} in the rearrangement.
We will refer to a rearrangement as localized if a small fraction of the atoms
participate in the rearrangement, and as
delocalized in the opposite limit.
The second index we define, $N_c$, 
is intended to characterize the number of atoms that move
simultaneously, i.e.~{\em cooperatively}. We will refer to a rearrangement as cooperative if most of the atoms
that participate in the rearrangement move simultaneously, and as uncooperative otherwise.

\begin{figure}
\centerline{\includegraphics{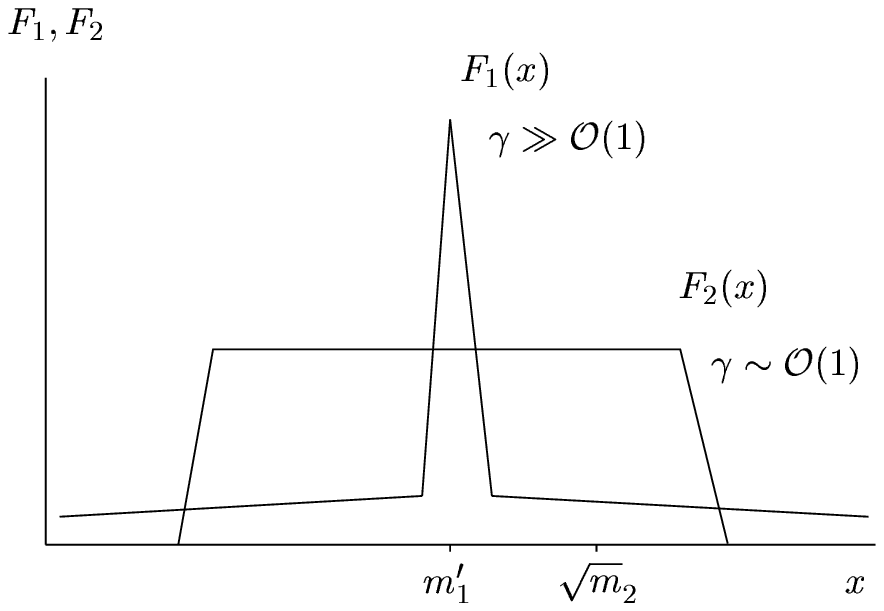}}
\caption{}
\label{fig:frequency}
\end{figure}

The $n$th moment about the mean for a data set $\{x_1, x_2, ... ,x_M \}$ is the expectation value of $(x_i - \langle x\rangle )^n$,
where $\langle x \rangle =\sum_{i=1}^M x_i /M$ and $M$ is the number of elements in the set.
Hence for the set $\{d\}$ defined above we define the moments, $m_n$, as
$m_n = \sum_{i=1}^N \left(d_i - \langle d \rangle \right)^n/N$.
The {\it kurtosis\/} of the set $\{d\}$ is then defined as the dimensionless ratio $\gamma(d) = m_4/(m_2)^2$,
and provides a measure of the shape of the frequency distribution function corresponding to $\{d\}$.
If only one of the atoms moves, or all atoms except one move by the same amount,
then $\gamma(d) = N$. Alternatively, if half the atoms move by the same amount whilst the others are stationary,
then $\gamma(d) = 1$. Hence, a distribution with a broad peak and rapidly decaying tails will have a small kurtosis,
$\gamma\sim{\cal O}(1)$, while a distribution with a sharp peak and slowly decaying tails will have a larger value
(Fig.~\ref{fig:frequency}).
The kurtosis can therefore identify distributions that contain large deviations from 
the average value. \cite{bulmer79}
For comparison, a Gaussian distribution has $\gamma=3$ and a uniform distribution has $\gamma=1.8$.

The above results show that the kurtosis $\gamma(d)$
can be used to quantify the degree of localization or delocalization of a given rearrangement.
However, it has the serious disadvantage that highly localized and delocalized mechanisms both have large values of
$\gamma(d)$. Since we
are interested in estimating the number of atoms that move relative to the number
with small or zero displacements,
a better approach is to use moments taken about the origin, rather than
about the mean, i.e.~$m'_n = \sum_i d_i^n/N$, following 
Stillinger and Weber.\cite{stillingerw83}. Note that while $m_1=0$,
the first moment $m'_1$ is the mean value.
We therefore estimate the number of atoms that participate in the
rearrangement, $N_p$, as
\begin{equation}
	N_p = \frac{N}{\gamma'(d)},
	\label{eq:np}
\end{equation}
where $\gamma'(d) = m'_4/\left(m'_2\right)^2$.
For the system with $N$ atoms, if only one atom moves $N_p=1$, while if $K$ atoms move
by the same amount, $N_p=K$.

A similar index to $N_p$ has been employed in previous work~\cite{wales94a,Wales03,ballbklpw96}
using only the displacements
between the two local minima, which corresponds to taking $N_f=2$ in Eq.~(\ref{eq:di}).
Using $d_i$ values based upon a sum of displacements that approximates the integrated path
length for atom $i$, rather than the overall displacement between the two minima,
better reflects the character of the rearrangement, as it can account for the nonlinearity of the pathway.
To describe this property more precisely we introduce a pathway nonlinearity index defined by $\alpha = (s - D)/s$,
where $D$ is the Euclidean distance between the endpoints,
\begin{equation}
	D =  \sqrt{\sum_{i=1}^{N} \left( {\bf r}_i(N_f)-{\bf r}_i(1) \right)^2}.
\end{equation}

We calculated the $\alpha$ values for a database of 31,342 single transition
state pathways of LJ$_{75}$ (hereafter referred to as the LJ$_{75}$ database).
The average value of $\alpha$ was $0.4$ with a standard deviation of $0.2$, and,
hence, there is a significant loss of information if $\gamma'$ is calculated only from
the endpoints using $N_f=2$. 
Comparison of the two indices for the LJ$_{75}$ database revealed many examples where neglect
of intermediate structures produces a misleading impression of the number of atoms that move.
The definition in equation~(\ref{eq:np}) is therefore suggested as an improvement on
previous indices of localization\cite{ballbklpw96,wales94a,Wales03,stillingerw83}.

\subsection{Cooperativity}
$N_p$ atoms can participate in a rearrangement according to
a continuous range of cooperativity. At one end of the scale
there are rearrangements where $N_p$ atoms all move simultaneously
[see Fig.~\ref{fig:displacement}(a)].
Although these paths exhibit the highest degree of correlated atomic motion they
do not usually pose a problem for double-ended 
transition state search algorithms.\cite{henkelmanjj00,trygubenkow04}
Linear interpolation between the two minima tends to generate initial guesses that lie close
to the true pathway, particularly if $\alpha \sim 0$. At the opposite extreme,
atoms can move almost one at a time, following a `domino' pattern
[see Fig.~\ref{fig:displacement}(b)].
Locating a transition state for such rearrangements may require a better initial guess,
since linear interpolation effectively assumes that all the coordinates change at the same rate.

The degree of correlation in the atomic displacements can be quantified
by considering the displacement `overlap'
\begin{equation}
	O_{k} = \sum_{j=2}^{N_f} O_{k}\left(j\right) = \sum_{j=2}^{N_f} {\rm min}\Bigl[ d_{c(1)}(j), d_{c(2)}(j), ..., d_{c(k)}(j)\Bigr],
	\label{eq:ok}
\end{equation}
where the index $k$ indicates that $O$ was calculated for $k$ atoms
numbered $c(1)$, $c(2)$,...,$c(k)$. ${\bf c}$ is a $k$-dimensional
vector that represents a particular choice of $k$ atoms from $N$, and hence
there are $C_N^k = N!/k!(N-k)!$ possible values of $O_{k}$.
The index $O$ can be thought of as a measure of how the displacements
of the atoms $c(1)$, $c(2)$, etc.~overlap along the pathway. For example, if two atoms move at different times then $O_2$ is
small for this pair because the minimum displacement in Eq.~(\ref{eq:ok}) is always small. However,
if both atoms move in the same region of the path then $O_2$ is larger.

\begin{figure}
\centerline{\includegraphics{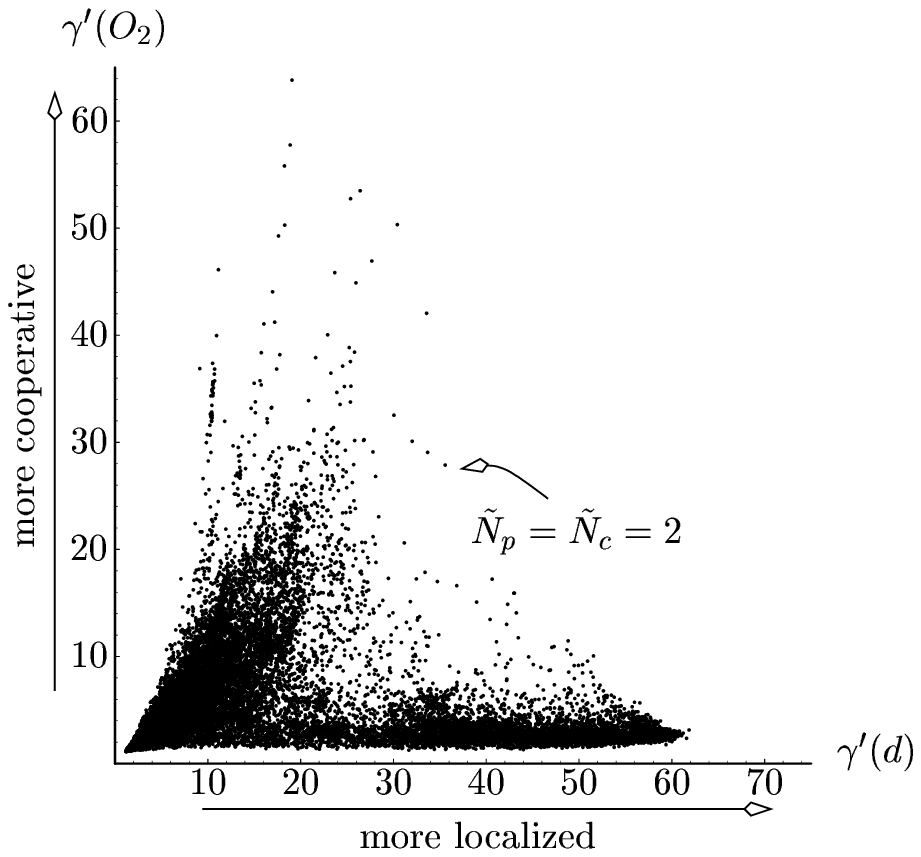}}
\caption{}
\label{fig:twooverlap}
\end{figure}

We now explain how the statistics of the overlaps, $O_k$, can be used to extract a measure of cooperativity
(Fig.~\ref{fig:twooverlap}). Suppose that $m$ atoms move simultaneously in a hypothetical rearrangement.
Then all the overlaps $O_k$ for $k > m$ will be relatively small, because one or more atoms are included
in the calculation whose motion is uncorrelated with the others. For overlaps $O_k$ with $k \leqslant m$ the set of $O_k$ for all
possible choices of $k$ atoms from $N$ will exhibit some large values and some small. The large values occur when all the chosen
atoms are members of the set that move cooperatively, while other choices give small values of $O_k$. Hence the kurtosis of the
set $\{O_k\}$, $\gamma'(O_k)$, calculated from moments taken about the origin,
will be large for $k \leqslant m$, and small for $k > m$. 
To obtain a measure of how many atoms move cooperatively we could therefore calculate $\gamma'(O_2)$, $\gamma'(O_3)$, etc.~and look
for the value of $k$ where $\gamma'(O_k)$ falls in magnitude.
However, to avoid an arbitrary cut-off, it is better to calculate
the kurtosis of the set $\{\gamma'(O_2)$, $\gamma'(O_3),...,\gamma'(O_k)\}$, or $\gamma'[\gamma'(O)]$ for short. There are $N-2$
members of this set, and by analogy with the definition of $N_p = N/\gamma'(d)$, we could define a cooperativity index
$N_c = (N-2)/\gamma'[\gamma'(O)] + 1$. Then, if $\gamma'(O_2)$ is large, and all the other $\gamma'(O_k)$ are small,
we obtain $\gamma'(\gamma'(O)) \sim N-2$ and $N_c \sim 2$, correctly reflecting the number of atoms that move together.

In practice, there are several problems with the above definition of $N_c$.
Calculating $N_c$ in this way
quickly becomes costly as the number of atoms and/or number of frames in the pathway increases,
because the number of elements in the set $\{O_k\}$ varies combinatorially with $k$.
Secondly, as $k$ approaches $N$ the distribution of all the possible values for $O_{k}$
becomes more and more uniform. Under these circumstances
deviations from the mean that are negligible in comparison with the overall displacement can produce
large kurtosises. Instead, we suggest a modified (and simpler) definition of $N_c$, which better satisfies our objectives.

We first define the overlap of atomic displacements in a different manner.
It can be seen from equation (\ref{eq:ok}) that the simultaneous
displacement of $l$ atoms is included in each set of overlaps $\{O_k\}$ with $k \leqslant l$.
For example, if three atoms move cooperatively then both the $\{O_2\}$ and $\{O_3\}$ sets
will include large elements corresponding to these contributions.
Another redundancy is present within $\{O_k\}$, since
values in this set are calculated for all possible subsets of $k$ atoms
and the displacement of each atom is therefore considered more than once.
However, we can avoid this redundancy
by defining a single $k-$overlap, rather than dealing with $C_N^k$ different values.

\begin{figure}
\centerline{\includegraphics{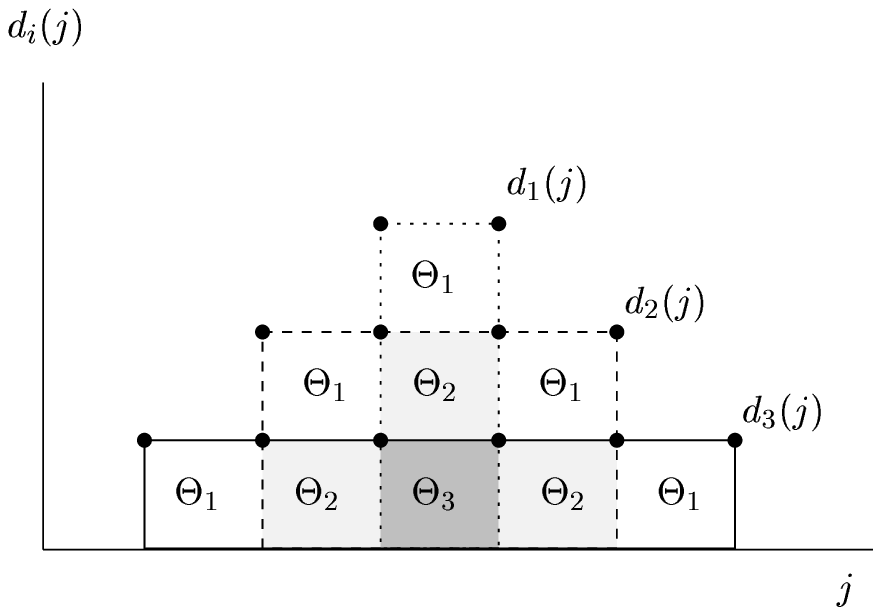}}
\caption{}
\label{fig:theta}
\end{figure}

Recall that $d_i(j)$ is the displacement of atom $i$ between frames $j-1$ and $j$. The ordering of the atoms
is arbitrary but remains the same for each frame number $j$. We now define $\Delta_i(j)$ as the displacement of
atom $i$ in frame $j$, where index $i$ numbers the atoms in frame $j$ in descending order, according to the magnitude of
$d_i(j)$, e.g.~atom $1$ in frame $2$ is now the atom with the maximum displacement between frames $1$ and $2$,
atom $2$ has the second largest displacement etc. As the ordering may vary from frame to frame, the atoms labelled $i$
in different frames can now be different. This relabelling greatly simplifies the notation we are about to introduce.
Consider the $k$-overlap defined as
\begin{equation}
	\Theta_k = \frac{1}{\Delta_{tot}}\sum_{j=2}^{N_f} \Bigl[ \Delta_k(j) - \Delta_{k+1}(j)\Bigr],
\end{equation}
where $k$ ranges from $1$ to $N$, $\Delta_{tot}=\sum_{j=2}^{N_f} \Delta_1(j)$ and $\Delta_{N+1}(j)$ is defined to be zero
for all $j$. A schematic illustration of this construct is presented in
Fig.~\ref{fig:theta}. For example,
if only two atoms move in the course of the rearrangement, and both are displaced by the same amount (which
may vary from frame to frame), the only non-zero overlap will be $\Theta_2$.

We can now define an index to quantify the
number of atoms that move cooperatively as
\begin{equation}
	N_c = \sum_{k=1}^{N} k \Theta_k.
	\label{eq:nc}
\end{equation}
If only one atom moves during the rearrangement then $N_c = 1$, while if
$K$ atoms displace cooperatively during the rearrangement then $N_c = K$. This definition
is independent of the total displacement, the integrated path length, and the number of atoms, which makes
it possible to compare $N_c$ indices calculated for different systems. 

\section{Results}
\begin{figure}
\centerline{\includegraphics{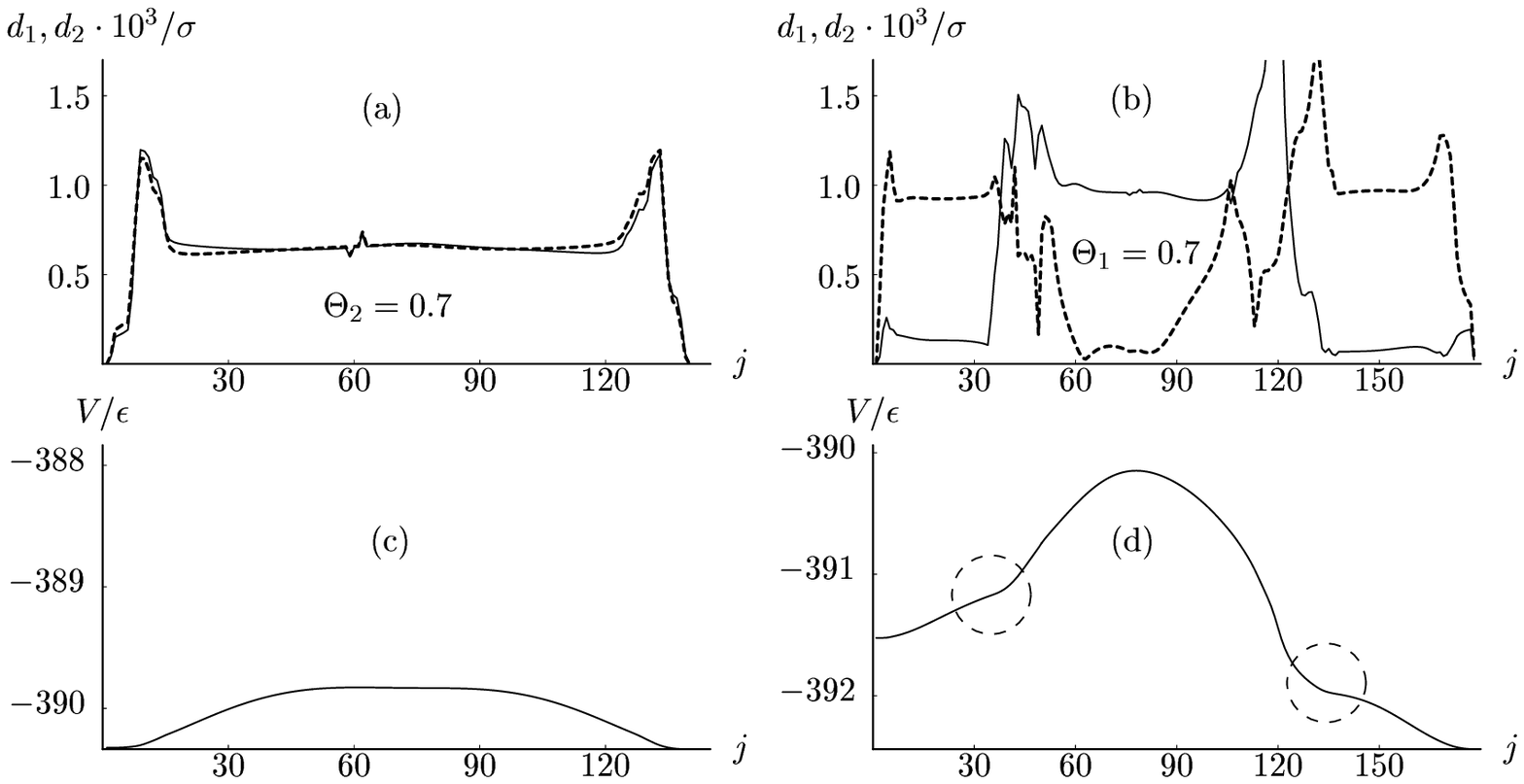}}
\caption{}
\label{fig:displacement}
\end{figure}

Fig.~\ref{fig:displacement} shows results for the most cooperative and uncooperative processes
we have found for the LJ$_{75}$ cluster that are localized mainly on two atoms.
In these calculations we have used the database of transition states that was found previously
as the result of a discrete path sampling calculation conducted for this system.\cite{Wales02,Wales04}
The cooperative rearrangement
[Fig.~\ref{fig:displacement}(a,c)] is the one with the maximum two-overlap $\Theta_2$. For this
pathway $\Theta_2=0.7$, $N_p=3.4$, and $N_c=7.7$. The values of $N_p$ and
$N_c$ both reflect the fact that the motion of the two atoms is accompanied by a slight distortion
of the cluster core. This example shows that while $N_p$ and $N_c$
allow us to quantify localization and cooperativity, and 
correctly reflect the number of atoms that participate and move cooperatively in ideal cases, there will
not generally be a simple correspondence between their values and the number of atoms that move.
This complication is due to the fact that
small displacements of atoms in the core will generally occur, no matter how localized the rearrangement is.
In addition, the data reduction performed in equations (\ref{eq:np}) and (\ref{eq:nc})
means that a range of pathways can give the same value for $N_p$ or $N_c$.
Since the size of the contribution from a large number of small displacements
depends on the shape of the displacement distribution function
the number of possibilities grows with the size of the index. 

The uncooperative rearrangement depicted in the Fig.~\ref{fig:displacement}(b,d) was harder to identify.
In principle we could have selected the pathway that maximizes $\Theta_1$
from all the rearrangements localized on two atoms. However, 
this approach picks out rearrangements localized on one atom, where distortion of the core
accounts for the value of $N_p>1$. Instead,
we first selected from all the rearrangements with $N_p < 4$
those where two atoms move by approximately the same amount, while
the displacement of any other atom is significantly smaller. These are the rearrangements
that maximize $1/ \gamma (\{d_1,d_2\})$, where $d_1$ and $d_2$ are the total displacements of the
two atoms that move the most. After this procedure we selected the rearrangement with the maximum
value of $\Theta_1$. Fig.~\ref{fig:displacement}(b) shows that this rearrangement
features the displacement of one atom at a time, and the atom that moves first also moves last.
For this pathway the values of $\Theta_1$, $N_p$ and $N_c$
are $0.7$, $3.8$ and $5.3$, respectively. Further illustrations and movies of the corresponding
rearrangements are available online.\cite{lj75pathways}

Fig.~\ref{fig:displacement} illustrates several general trends that we have observed for cluster rearrangements.
Firstly, we have found that the barrier height is smaller for the cooperative rearrangements
[Fig.~\ref{fig:displacement}(c,d)].
Usually atoms that move cooperatively are neighbours. Rearrangements generally involve
a change of the environment for the atoms that move.
Cooperative motion can reduce this perturbation since for any of the participants
the local environment is partly preserved because it moves with the atom in question. Flatter points on the energy profile
[circled in Fig.~\ref{fig:displacement}(d)] usually signify a change in the mechanism, i.e.~one
group of atoms stops moving and another group starts. By comparing (b) and (d) in Fig.~\ref{fig:displacement}
we conclude that flatter points on the energy profile 
correlate with the most cooperative parts of this rearrangement.

\begin{figure}
\centerline{\includegraphics{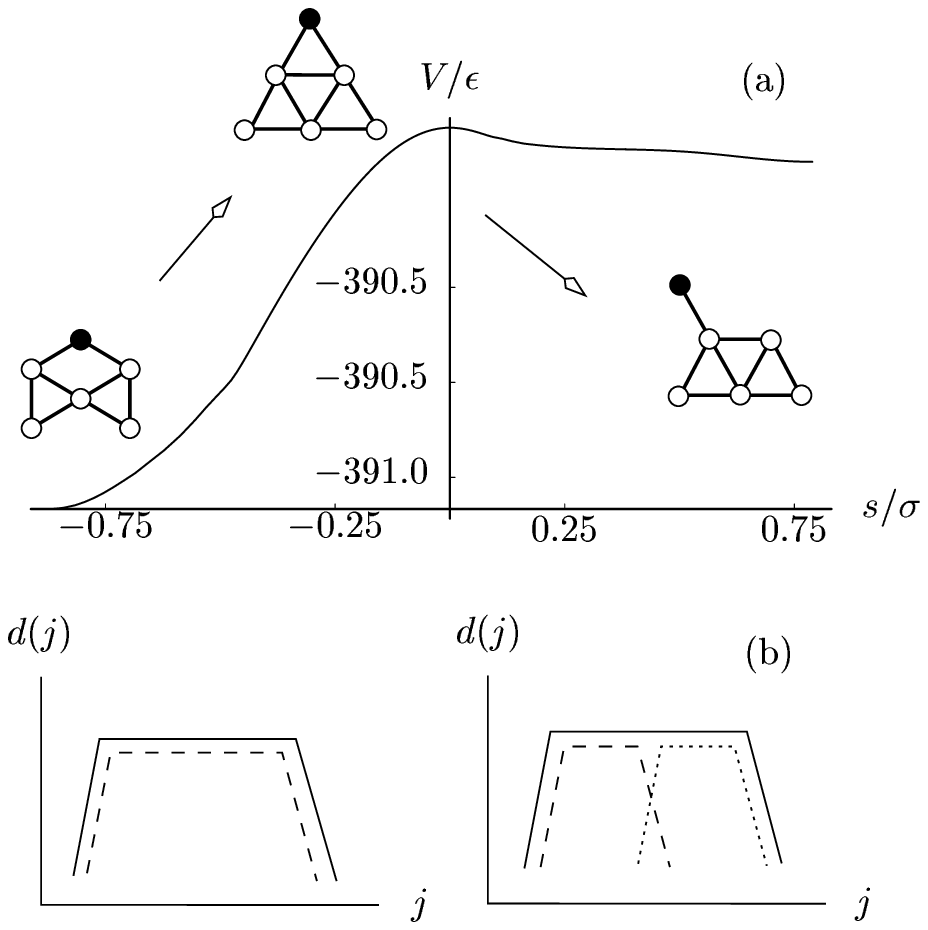}}
\caption{}
\label{fig:limitations}
\end{figure}

A simple correlation between barrier heights and $N_p$ and $N_c$ does not seem to exist.
The barrier height is not a function of 
cooperativity alone, but also of the energetics of the participating atoms.
The way the $N_p$ and $N_c$
indices have been defined can make them insensitive to details of the rearrangements that
will affect the energetics. For instance, neither index depends on the location
of the participating atoms or the directionality of their motion.
In most cases cooperatively moving particles are adjacent, i.e.~localized in space; however,
long-distance correlations of atomic displacements also occur. One such case is depicted in Fig.~\ref{fig:limitations} (a).
This path is nearly symmetric 
with respect to the integrated path length ($\pi = 0.01$, see Ref.~\citen{trygubenkow04}), but is very asymmetric
with respect to the uphill and downhill barrier heights ($\beta = 0.91$, see Ref.~\citen{trygubenkow04}).
This rearrangement has $N_p = N_c = 10$.
Interestingly, $N_p$ and $N_c$ calculated separately for both sides of the pathway
are very similar, i.e.~the two steepest-descent pathways cannot be distinguished
using these indices. Close inspection of this rearrangement reveals that one side of the pathway involves
the rearrangement of two atomic triplets that share a vertex, while the other side
involves the drift of all five atoms on the surface of the cluster [see the insets in Fig.~\ref{fig:limitations} (a)].
Although $N_c$ does not distinguish these cases, the motion in the second side of the path
is more cooperative. 
The participating atoms move together, which results in a significantly lower downhill barrier.

\begin{figure}
\centerline{\includegraphics{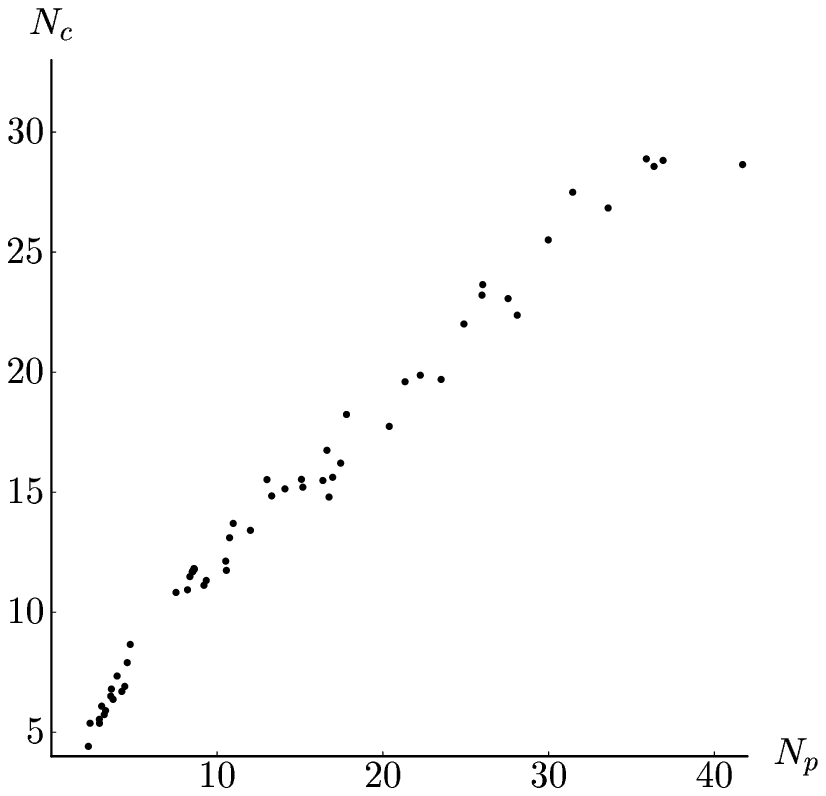}}
\caption{}
\label{fig:coop}
\end{figure}

$N_p$ and $N_c$ also describe properties of the whole pathway. A
significant number of pathways that we observed were rather non-uniform, i.e.~very
cooperative phases alternated with uncooperative ones. To distinguish such pathways
in the LJ$_{75}$ database we calculated a set $\{N_p\}$ containing
$N_p$'s evaluated for each pair of adjacent frames. Then a selection of pathways was made with
$m_2/(m'_1)^2 < 0.01$, where $m_2/(m'_1)^2$ is the moment ratio evaluated for the set $\{N_p\}$.
While this procedure ensured that $N_p$ corresponds closely
to the number of atoms that moves between any two snapshots of the rearrangement, it
did not distinguish cases where different atoms contribute to the value of $N_p$ in
different frames [see Fig.~\ref{fig:limitations}(b)].
The average uphill and downhill barriers for this subset of rearrangements
are $100$ times smaller than the average barriers for the complete LJ$_{75}$ database (Table \ref{tab:barriers}).
Fig.~\ref{fig:coop} shows that $N_p$ and $N_c$ calculated for these rearrangements
are highly correlated and span a range of values,
implying that widely different pathways are represented. 
Finally, all the selected pathways are an order of magnitude
shorter than the average path length for the whole database,
even though this database contains many short rearrangements localized on one atom.
% Database:
% Rf min longer = 1.069546341
% Rf ts longer  = 1.076900327
% Rf ts shorter = 1.034660549
% Rf min shorter= 1.485803923
% 
% Subset:
% Rf min longer = 1.943414285
% Rf ts longer  = 1.872354106
% Rf ts shorter = 1.445891694
% Rf min shorter= 0.952178191 

Fig.~\ref{fig:ntildes} shows the values of the $N_p$ and $N_c$ indices plotted against
each other for pathway databases calculated for LJ$_{13}$, LJ$_{75}$ and BLJ$_{256}$. The two databases for BLJ$_{256}$
labelled as $1$ and $12$ are taken from Ref.~\citen{middletonw01} and correspond to 
databases BLJ1 and BLJ12 in that paper. BLJ1 and BLJ12 were obtained using two
different sampling schemes intended to provide an overview of a wide range of configuration space and a
thorough probe of a smaller region, respectively. These databases were constructed by
systematic exploration of the PES, and we refer the reader to the original work for further details.\cite{middletonw01}
Each BLJ$_{256}$ database contains $10,000$ transition states. The LJ$_{13}$ and LJ$_{75}$ databases were obtained
in discrete path sampling (DPS) studies~\cite{Wales02,Wales04}
and contain $28,756$ and $31,342$ transition states, respectively. Fig.~\ref{fig:ntildes} is a density plot
where darker shading signifies a higher concentration of data points. The outlying points
are connected by a solid line to define the area in which all the points lie.
Fig.~\ref{fig:ntildes} shows that as $N_p$ grows the allowed range of $N_c$ increases,
especially for LJ$_{13}$. For the LJ$_{75}$ database rearrangements
with $N_c > N/2$ appear to be very rare or poorly sampled. Fig.~\ref{fig:ntildes} also shows
that for all these systems rearrangements localized on two or three atoms dominate.
This result may be an intrinsic property. However, it may also be due to the geometric perturbation scheme
used in producing the starting points for the transition state searches employed in generating these databases.
For databases LJ$_{13}$, BLJ1 and BLJ12 there are significantly more rearrangements with larger values of
$N_p$ and $N_c$
compared to LJ$_{75}$, which suggests that the abundance of very localized rearrangements for clusters may be
a surface effect.

\begin{figure}
\centerline{\includegraphics{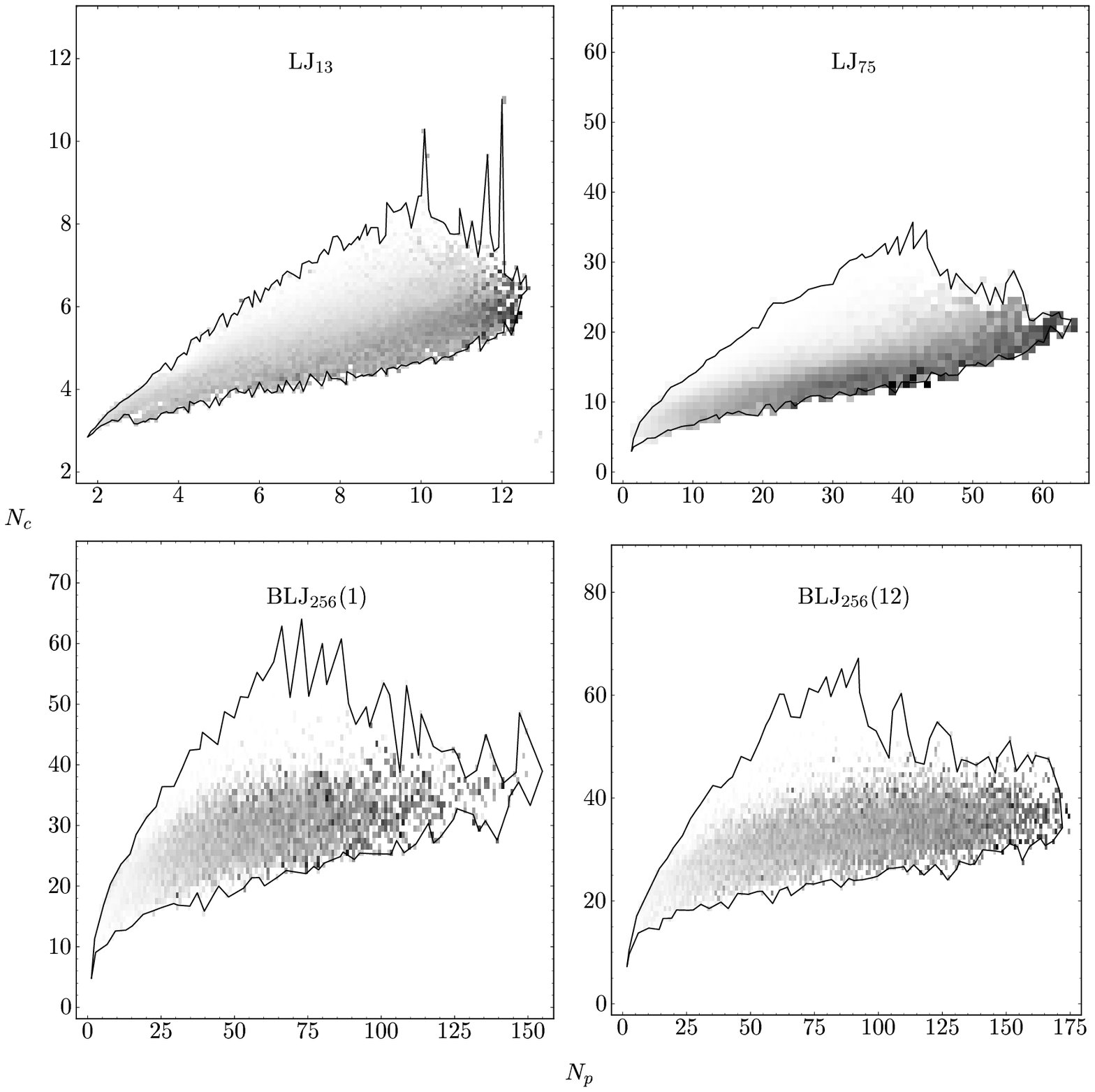}}
\caption{}
\label{fig:barriers}
\end{figure}

Fig.~\ref{fig:barriers} depicts the average barrier as a function of the participation and cooperativity indices.
$N_p$ and $N_c$ were calculated separately for both sides of the pathway and the corresponding
barriers (uphill or downhill) were averaged to produce a density plot of barrier height. This figure illustrates
that for each system cooperative rearrangements 
have the lowest barriers, irrespective of the value of $N_p$.
For clusters, cooperative rearrangements have lower barriers than uncooperative rearrangements
with $N_p$ as small as $1-3$, while for bulk
barriers corresponding to rearrangements with low $N_p$ become comparable to these for
very cooperative rearrangements with high $N_p$.

\begin{figure}
\centerline{\includegraphics{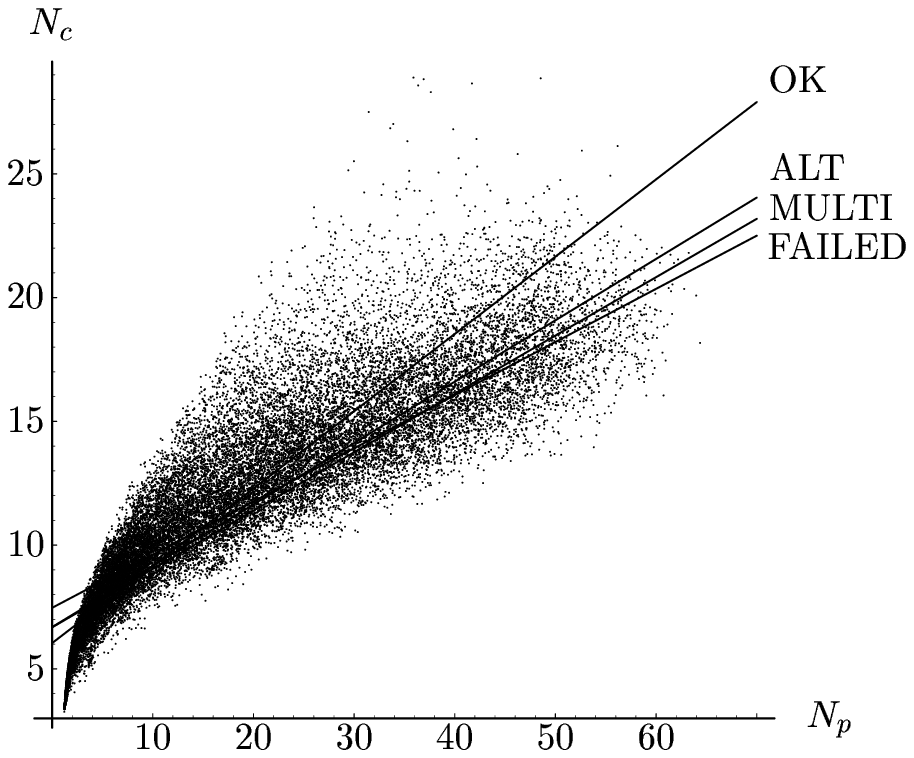}}
\caption{}
\label{fig:dnebproblems}
\end{figure}

In further computational experiments we find that attempts to connect the endpoints of uncooperative pathways
using the algorithm described in Ref.~\citen{trygubenkow04} either required more images and iterations
or converged to an alternative pathway. In some cases additional difficulties arose, such as
convergence to a higher index saddle instead of a transition
state, which can happen if the linear interpolation guess
conserves a symmetry plane. Fig.~\ref{fig:dnebproblems} shows
$N_c$ calculated from Eq.~(\ref{eq:nc}) plotted against
$N_p$ for the LJ$_{75}$ pathway database. Knowing the integrated
path length, $s$, for each pathway we started doubly nudged elastic band
calculations with three images per unit of distance and $30$ iterations per image.
Most of the points that correspond to runs that failed or converged to an alternative pathways
are concentrated in the region of small values for $N_c/N_p$.

\begin{figure}
\centerline{\includegraphics{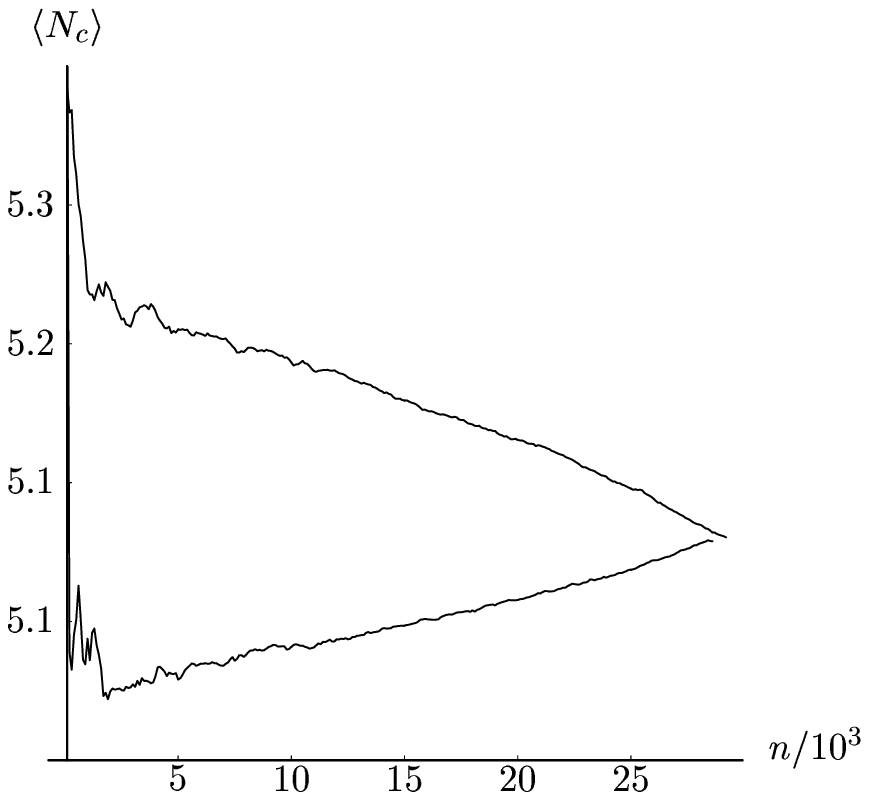}}
\caption{}
\label{fig:sampling}
\end{figure}

As can be seen from Fig.~\ref{fig:ntildes}, the LJ$_{13}$ database contains significantly more
pathways with large values of $N_p$ and $N_c$ compared
to LJ$_{75}$, where most of the rearrangements are localized on two atoms. The fact that the LJ$_{13}$ database
is almost exhaustive then suggests that localized rearrangements either
start to dominate as the system size increases or that the sampling scheme used for LJ$_{75}$
was biased towards such mechanisms.
Systematic sampling of the configuration space for stationary points often employs perturbations of every degree of
freedom followed by minimization.\cite{walesd97} The LJ$_{13}$ database was obtained in this fashion,
while the LJ$_{75}$ database was generated during the discrete path sampling (DPS) approach.\cite{Wales02}
In this procedure discrete paths are perturbed by replacing local minima with structures obtained after perturbing
all the coordinates and minimizing. To investigate whether the perturbation scheme can affect
the resulting database of stationary points in more detail we consider the case of LJ$_{13}$, since
nearly all the transition states are known. Fig.~\ref{fig:sampling} presents the results of two
independent runs aimed at locating most of
the transition states for this system. 
Every cycle a perturbation was applied to a randomly selected transition state from the database
and the resulting geometry was used as a starting point for a new transition state search using
eigenvector-following.\cite{wales94a,walesw96,munrow99,kumedamw01,jensen95} Only distinct permutation-inversion isomers
were saved. In the first run (bottom curve) every degree of freedom was perturbed by $0.4 x$, where $x$ is a random number
in the interval $[-1,1]$.\cite{walesd97} For the second run (top curve) we introduced a perturbation scheme including correlation.
$2 \leqslant K \leqslant N/2$ atoms out of $N$ were displaced by a vector $0.4(x_1, x_2, x_3)$, where
the components $x_1$, $x_2$ and $x_3$ are again random numbers 
drawn from $[-1,1]$. The $K$ atoms to be displaced were selected based
on their relative positions in the cluster. One atom was first selected at random, while the remaining
$K-1$ were chosen to be its closest neighbours. The top curve was generated from a run with $K=6$.
Both runs required approximately the same time to produce two nearly identical
databases, each containing about $29,000$ pathways. However, as can be seen from Fig.~\ref{fig:ntildes}, random perturbation
of all the degrees of freedom results in uncooperative rearrangements being found first, while employing correlated perturbations
has the opposite effect.

\section{Conclusions}
The most important result of this work is probably the introduction of an index to quantify the
cooperativity of atomic rearrangements. With this new measure
it becomes possible to correlate cooperativity and barrier heights, and to show that cooperative
rearrangements generally have lower barriers and shorter path lengths.
We hope that these results will shed new light on relaxation mechanisms in complex systems,
such as glasses and biomolecules, in future applications.
For example, in a peptide or protein a large geometrical change can result from a rearrangement
that could be described in terms of a single dihedral angle.
In glasses and supercooled liquids an important research goal is to understand
how observed dynamical properties, such as atomic diffusion and 
correlation functions,\cite{middletonw01,vogeldhg04,GebremichaelVG04,Jaind04}
are related to features of the underlying PES.
The classification of elementary rearrangements as 
`cage-breaking' or `cage-preserving',\cite{middletonw01,MiddletonW03}
and the emergence of structures such as 
`megabasins'\cite{middletonw01,DoliwaH03,MiddletonW03,SaksaengwijitDH03} 
can now be investigated more precisely in terms of localization and cooperativity.

We have also demonstrated that cooperative rearrangements are relatively easy to characterize using double-ended transition
state searching algorithms, since linear interpolation produces an effective initial guess. Uncooperative rearrangements
are usually harder to find using such methods,
and alternative initial guesses may be helpful in these cases.

Single-ended transition state searching has been used both in conjunction with double-ended
methods, and as a way to sample potential energy surfaces for
stationary points. Stationary point databases constructed using random perturbations followed by quenching
are likely to be biased towards uncooperative rearrangements. We have therefore outlined 
a strategy for generating initial guesses appropriate to single-ended transition state searching algorithms,
which instead favours cooperative rearrangements. This approach also includes a parameter that is likely to influence
the degree of localization.

\section{Acknowledgements}
S.A.T.~is a Cambridge Commonwealth Trust/Cambridge Overseas Trust scholar. Most of the calculations
were performed using computational facilities funded by the Isaac Newton Trust. We are grateful to
T.~V.~Bogdan for useful discussions and to Dr.~T.~F.~Middleton
for providing transition state databases for the binary Lennard-Jones system. 
Financial support from Darwin College, Cambridge
is also gratefully acknowledged.

\bibliographystyle{008439JCP}
\bibliography{008439JCP}
\clearpage

\section{Tables}

\begin{table}[h]
\caption{Average uphill and downhill barriers, average integrated path length
for the LJ$_{75}$ rearrangement pathway database. Values are given
for the whole database containing $31,342$ paths,
and for a subset containing the $57$ most cooperative paths. The units of energy and distance are
$\epsilon$ and $\sigma$, respectively. }
\label{tab:barriers}
\begin{center}
\begin{tabular}{llcc}
\hline
\hline
                       & All      & Cooperative     \\
\hline
   Uphill barrier      & 3.03     & 0.06            \\
 Downhill barrier      & 0.97     & 0.03            \\
 Path length           & 3.08     & 0.58            \\
\hline
\hline

\end{tabular}
\end{center}
\end{table}

\clearpage

\section{Figure Captions}
\begin{figure}
\centerline{\includegraphics{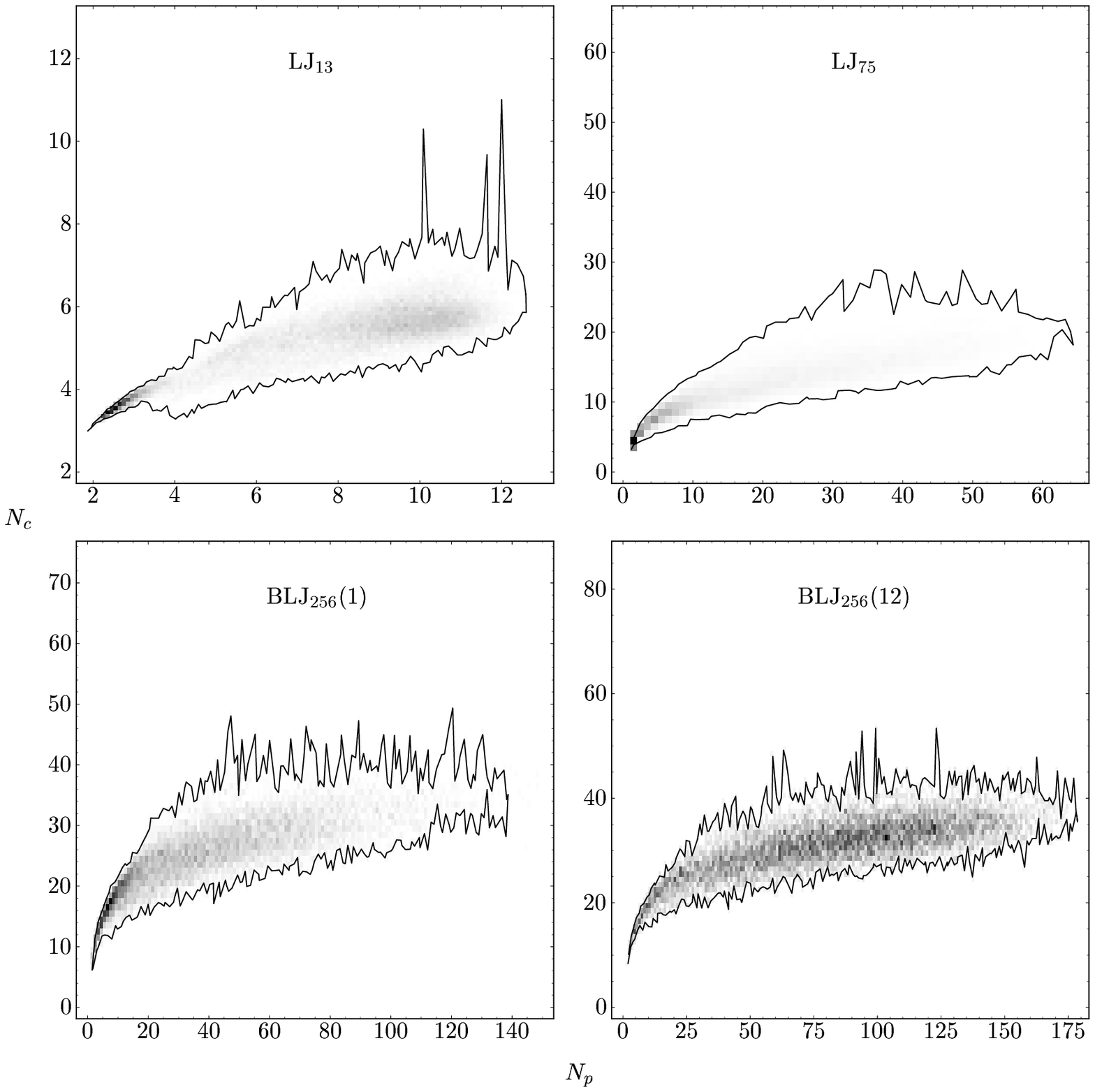}}
\caption{}
\label{fig:ntildes}
\end{figure}

\begin{enumerate}
\item
Two frequency distribution
functions $F_1$ and $F_2$ of a continuous variable $x$ are contrasted.
Both functions have the same average $m'_1$ and standard deviation
$\sqrt m_2$. However, due to the long tails, $F_1$ has a significantly larger
fourth moment $m_4$ and, hence, a larger kurtosis, $\gamma$.
\item
Comparison of cooperative (a) and uncooperative (b) rearrangements of the LJ$_{75}$ cluster, for mechanisms that are
localized mainly on two atoms. The displacement $d$ as a function of the frame number $j$
is shown for the two atoms that move the most. Panels (c) and (d) illustrate the potential energy $V$
as a function of frame number $j$ for the rearrangements in (a) and (b), respectively.
Dashed circles indicate flatter parts of the energy profile, which correspond to the most cooperative regions
of the pathway.
\item
$\gamma'(O_2)$ plotted against $\gamma'(d)$ for the LJ$_{75}$ pathway database.
The figure shows how $\gamma'(O_2)$ can discriminate between rearrangements that have similar values of $\gamma'(d)$ but different
cooperativity. The data point for the most cooperative rearrangement
localized on two atoms has $N_p=N_c=2$. 
\item
The $\Theta$ indices for a hypothetical rearrangement localized on three atoms. For each of these atoms the displacement $d$
as a function of frame number $j$ is shown. The $d_i$ values
in successive frames are connected with dotted ($d_1$), dashed ($d_2$) and solid ($d_3$) lines.
The corresponding contributions to $\Theta_1$, $\Theta_2$, and $\Theta_3$ are shown as shaded squares and are labelled accordingly.
If the remaining $N-3$ atoms do not participate, and the area of one square is $S$, the only non-zero overlaps
will be $\Theta_1$, $\Theta_2$, and $\Theta_3$ with values $5/9$, $3/9$ and $1/9$, respectively.
\item
Two limitations of the cooperativity index $N_c$. (a) $N_c$ is not sensitive
to the spatial positions of the cooperatively moving atoms, nor to the directionality of their motion.
The energy profile is depicted for a rearrangement of the LJ$_{75}$ cluster, which is very asymmetric with respect
to barrier height but has similar integrated path lengths on either side of the transition state.
The cooperativity index $N_c$
evaluated separately for the two sides is about $10$ in both cases.
The motion of the five atoms that displace the most is shown schematically relative to a reference atom (black).
(b) The displacement of two (left) and three (right) atoms $d$ as a function of the frame number $j$ is shown schematically
for a hypothetical pathway.
The rearrangement on the left is more cooperative because two atoms move together over a longer region of the path.
However, the current definition of $N_c$ does not distinguish between these two cases.
\item
$N_c$ as a function of $N_p$ calculated for the $57$ most cooperative
rearrangements from the LJ$_{75}$ pathway database.
\item
$N_c$ as a function of $N_p$ calculated from four pathway databases
for LJ$_{13}$, LJ$_{75}$ and BLJ$_{256}$ systems. Due to the large number of data points we employ a density plot,
where the darkest shading corresponds to the highest concentration of points.
Outlying points are connected to illustrate the boundaries of the data area. The two BLJ databases are taken
from Ref.~\citen{middletonw01}
\item
Average barrier as a function of $N_p$ and $N_c$ calculated for the same 
LJ$_{13}$, LJ$_{75}$ and BLJ$_{256}$ pathway databases used in Fig.~\ref{fig:ntildes}.
The indices were calculated separately for the two sides of each path.
In this case the darkest shading corresponds to the highest barriers.
\item
%$N_c$ as a function of $N_p$ calculated for the LJ$_{75}$ pathway database.
%For each pathway we conducted DNEB calculations~\cite{trygubenkow04} assuming prior knowledge of the path. Every
%DNEB calculation employed $3s$ images and $90s$ iterations in each case, where $s$ is the integrated path length.
%Out of $31,342$ DNEB runs $25,158$ yielded a connected pathway while the
%rest did not (FAILED, points shown in red). Connected pathways are classified further as
%one-step pathways involving the correct transition state (OK, black) or an alternative transition state (ALT, green),
%or as multi-step pathways (MULTI, blue), which involve more than one transition state.
%For each set of data points best fit straight lines obtained from linear regression are also shown and labelled appropriately.
$N_c$ as a function of $N_p$ calculated for the LJ$_{75}$ pathway database.
For each pathway we conducted DNEB calculations~\cite{trygubenkow04} assuming prior knowledge of the path. Every
DNEB calculation employed $3s$ images and $90s$ iterations in each case, where $s$ is the integrated path length.
Out of $31,342$ DNEB runs $25,158$ yielded a connected pathway while the
rest did not (FAILED). Connected pathways are classified further as
one-step pathways involving the correct transition state (OK) or an alternative transition state (ALT),
or as multi-step pathways (MULTI), which involve more than one transition state.
For each set of data points best fit straight lines obtained from linear regression are shown and labelled appropriately.
\item
The average value of $N_c$ for LJ$_{13}$ pathway databases as new paths are added.
6-atom correlated perturbations (top curve) and random perturbations of every degree of freedom (bottom curve)
were used to produce starting points for refinement by
eigenvector-following.\cite{wales94a,walesw96,munrow99,kumedamw01,jensen95} Average values were calculated
every time $100$ new pathways were added. 
\end{enumerate}
\clearpage

\clearpage

\clearpage

\clearpage

\clearpage

\clearpage

\clearpage

\clearpage

\clearpage

\clearpage

\clearpage

\end{document}